# The fractal dimension of music: Melodic contours and time series of pitch.

Maria H. Niklasson[1,*] and Gunnar A. Niklasson[2]

[1] Cathedral School, SE-75375 Uppsala, Sweden and Betel Music Institute, Campus Bromma, Åkeshovsvägen 29, SE-16839, Bromma, Sweden

[2] Department of Engineering Sciences, Solid State Physics, The Ångström Laboratory, Uppsala University, P.O. Box 534, SE-75121 Uppsala, Sweden

**Abstract**

We analyze the fractal dimension of melodic contours and pitch time series of classical music and folk music tunes. The fractal dimensions obtained from box counting and detrended fluctuation analysis show significant differences. They are ascribed to the low accuracy of box counting for dimensions close to two as well as to a possible bias because the pitches in the time series are connected by lines to obtain the melodic contour used in the box counting analysis. We observe a tendency that folk music exhibits lower fractal dimensions than classical music, but further studies are needed in order to assess cutoff effects in the comparatively short folk music tunes. We conclude that detrended fluctuation analysis is the preferable method for fractal analysis of music, and this verifies previous studies of analysis of short time series.

[*] Present address: Royal College of Music, SE-11591, Stockholm. Sweden

# 1. Introduction

Since the pioneering work of Mandelbrot [1] fractal geometry has been widely used in the characterization of structures in the natural world as well as in a variety of scientific fields. It was first noticed by Voss and Clarke [2,3] that the power spectral density of audio recordings show major similarities to 1/f noise. This can be understood in terms of fractal, or more precisely self-affine, structural properties of music, which may, at least to a first approximation, be modeled by fractional Brownian motion [4,5]. The discovery of fractal structure in music was popularized by Gardner [6]. As pointed out already in the early work [3,4] such analyses are important for the development of algorithmic composition [7], i.e. computer generated music, which is often used as a tool for developing novel ideas in musical composition.

The statistical properties of music can be analyzed using different methodologies. Measurements of the spectral density of audio power and frequency fluctuations in audio recordings is a versatile and fast method [2,3,8]. However, the results of such a procedure will depend both on the musical composition and how it is performed. In the present paper we take another approach and analyze the actual musical scores, in order to focus on the intention of the composer. It seems that Hsü and Hsü [9] made the first attempt to analyze fractal properties of musical scores. They defined a fractal dimension from the distribution of melodic intervals. However, this concept is suspect and has subsequently been strongly criticized [10]. A conceptually much better method is to digitize a musical score to create a time series of the pitch, a so called "melodic contour" or "music walk" [5].

In addition, different methods to determine the fractal dimension of music have been employed and the obtained fractal dimensions sometimes display significant differences. Box counting [11] is a well-known fractal analysis method which has been used in early work [12]. Fourier transformation has been used to obtain a power spectral density from the melodic contour [5] and from measures of rhythm [13]. Methods based on the variance of the audio amplitude [8] or the root-mean square (RMS) fluctuation [5] of the melodic contour have also been used to determine the fractal dimension, $D$. In the case of fractional Brownian motion the square root of the variance (i.e. the RMS fluctuation) scales as a power of the length of the considered time interval [4]. The scaling exponent, called the Hurst exponent, is given by the relation $H=2-D$ [4]. The RMS analysis can be improved by removing trends from the data and the resulting Detrended Fluctuation Analysis (DFA) [14,15], was recently used to analyze music [16]. It should also be noted that a variety of more complex methods have been used to study scaling in music, ranging from multifractal analysis [17-19], concepts from chaos theory [20] and a variety of other metrics [21] to determinations of special note patterns or fractal generators responsible for the scaling [22,23].

Recently we showed [16] that the distribution of melodic intervals, i.e. the increments in pitch between successive time steps in the melodic contour is strongly non-Gaussian and can to a fair approximation be fitted by Levy-stable distributions [24]. This suggests that a basic theme in the structure of music can be described as a Levy motion [25,26]. This finding is of potential importance for the analysis of the fractal structure of music since it has been claimed that methods based on the variance are not reliable in cases of non-Gaussian long-tailed distribution of increments [26]. Although many methods have been used to study fractal structures in music a coherent picture has not yet emerged. There still exist considerable uncertainties regarding the applicability and accuracy of the methods used to extract even the most basic parameter, i.e. the fractal dimension $D$. In the present paper we compare different methods to calculate the fractal dimension from the melodic contour of a number of classical pieces and folk tunes. We discuss the possible reasons for differences in the obtained numerical values and find a tendency that classical works exhibit a higher fractal dimension than folk music.

## 2. Fractal analysis methods

The music scores were digitized manually to obtain a time series of pitches, a so called melodic contour. Each note in the script was given an integer number characterizing the pitch and a number of time steps characterizing the duration of the note. The pitch is defined according to the twelve tone scale of Western music, in which each octave, which ranges from frequency $f$ to $2f$, is divided into twelve intervals between tones, each tone being associated with a pitch value. The pitch is related to the logarithm of the frequency of the tone and is represented by an integer. The frequencies in an octave are related to the pitches $i$ by [5]

$$\frac{f_i}{f_0} = 2^{i/12} \qquad (1)$$

where $f_0$ is the frequency of the base note in an octave. Since musical pieces span over several octaves, we number the pitches starting from the lowest note in the score. The shortest note in the score was taken as the fundamental time step. Hence a note may be assigned to one or more time steps, depending on its duration. Pauses were assigned an appropriate number of time steps but no pitch, hence they were treated as interruptions of the music. Figure 1 shows two examples of melodic contours, specifically for a classical work, Bach's Concerto for two violins in D minor, movement 2, and for a folk music tune, "Leksand gift tune". It should be noted that folk tunes generally are much shorter than classical ones, as seen from the different scales on the horizontal axes.

Box counting [11] is a method specially developed for self-similar fractals, but it can also be applied to self-affine surfaces and profiles [27], like melodic contours. An image of the structure is covered by boxes of size $L$ and the number of boxes, $N$, covering the contour is determined as a function of box size. For a fractal structure, [11,27]

$$N(L) \sim L^{-D} \qquad (2)$$

In order to avoid crossover effects between "local" and "global" dimensions, it is essential that the units in the vertical direction (pitch) and in the horizontal direction (time) are chosen so that the image to be analyzed exhibits an overall square shape [27,28]. We have analyzed images of melodic contours, like those in figure 1, by the box counting routine in the image analysis program ImageJ [29].

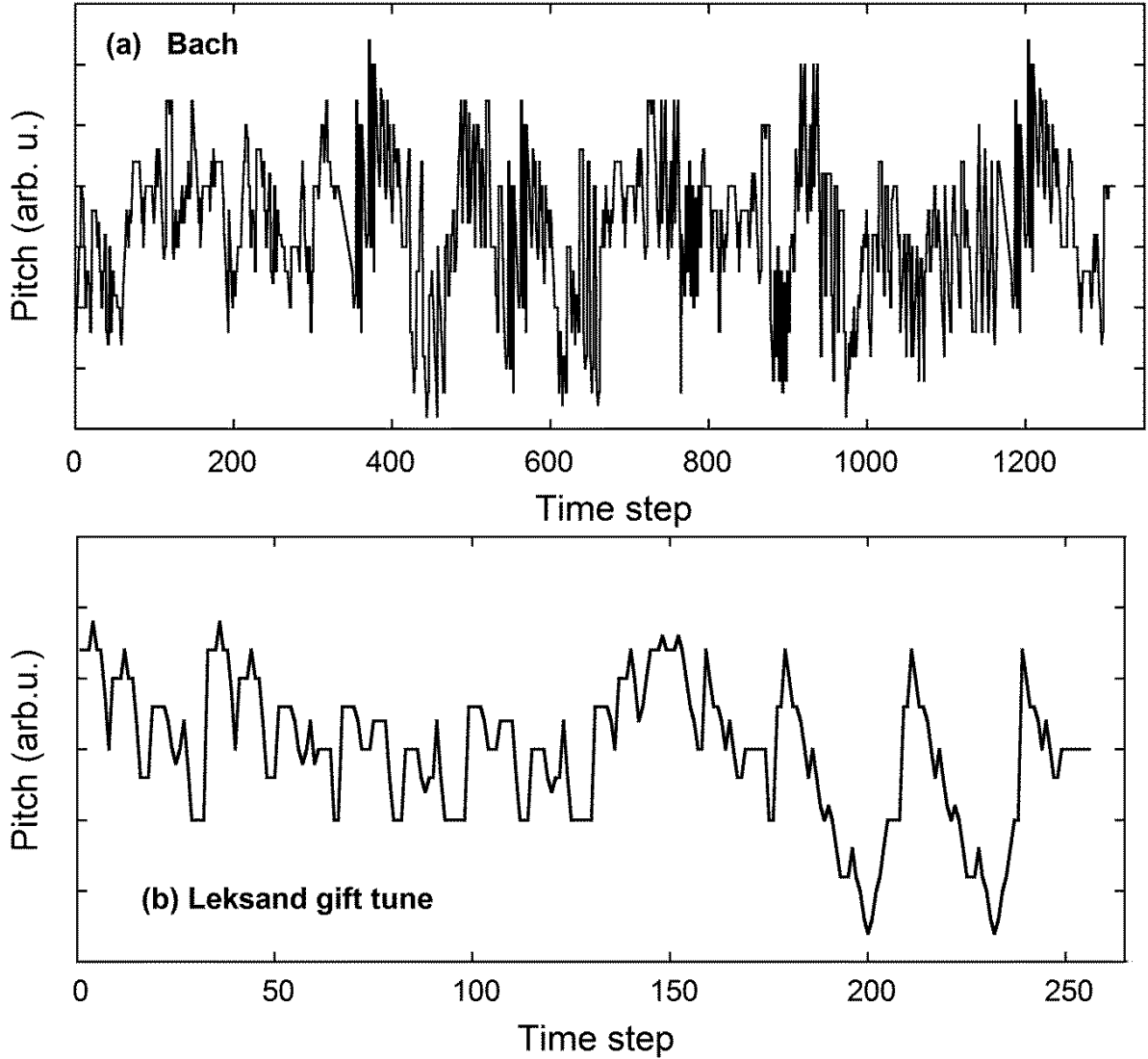

Fig. 1. Melodic contour showing pitch as a function of time step for (a) Bach's Concerto for two violins in D minor, movement 2, and (b) Leksand gift tune. The length of the vertical axis is 35 pitch units in (a) and 30 pitch units in (b).

Time series of pitch, like those of Figure 1, were considered analogous to random walks (a "music walk"), in order to apply RMS analysis. In this method the scaling of the RMS fluctuations was obtained by dividing the time series into non-overlapping subseries ("bins") of length $L$ and equal number of data points, $n$, and using the relation [27]

$$(<\frac{1}{n}\Sigma_n [h(t) - \bar{h}]^2 >)^{1/2} \sim L^H \, , \qquad (3)$$

where $h(t)$ is the pitch, $\bar{h}$ its average, the sum is over all the data in a bin and the brackets denote an average over all bins of size $L$. It is very important to decouple the analysis of a fluctuating signal from any underlying trend in the data. Detrended Fluctuation Analysis (DFA) [14,15] is a well-established method for advanced analysis of Hurst exponents. It should be realized that the pitch series *h(t)* corresponds to the cumulative time series of the DFA method [14,15]. In the DFA method the trend of the cumulative time series is determined by least-squares fitting of a polynomial to the data in each bin. In the present study we use linear and quadratic polynomials as trend curves. The RMS fluctuations of the detrended time series is calculated by replacing the average $\bar{h}$ in eq. (3) by the polynomial trend curve. Calculations are performed for each bin and averaged over all bins of the same size, in analogy to eq. (3). In the scaling range the DFA estimator, which is proportional to the RMS fluctuation, is $\sim L^H$ [14,15,30]. We have determined the Hurst exponent by the DFA method, using a Matlab function due to Weron [30,31]. The algorithm was modified to encompass the RMS method (DFA0), linear trend subtraction (DFA1) as well as quadratic trend subtraction (DFA2).

The accuracy of the fractal analysis is affected by crossovers between the fractal scaling range (*D=2-H*) in eq. (2) and (3) and a non-fractal range (*D=2*) at large *L*. This is because the behavior for long-time intervals is affected by the fact that we have a finite set of data with a finite span of pitch values. The DFA estimator cannot increase above the maximum given by the range of the data and this can lead to a crossover which is different for each data set. An advantage with detrending is that it may shift the crossover and extend the fractal scaling range towards larger *L*, however sometimes at the expense of introducing another crossover at small *L* [15].

The above computations were compared to alternative algorithms from the work of Russ [32]. His Kolmogorov method is very similar to box counting and gives slightly larger (0.05-0.1) values of *D* than found with our method. In addition, the RMS fluctuation routine of Russ was found to be in good agreement with our DFA0 calculations. Another method to determine H is the so-called rescaled range analysis, which was carried out by the R/S routine in the software SELFIS [33]. However, we found that this method exhibits ill-defined crossovers, which makes results very uncertain.

## 3. Results and discussion

Figure 2 shows box counting results for the two musical contours depicted in figure 1. The data follow closely a straight line according to eq. (2) with crossover effects at small box sizes, where the one-dimensional lines making up the melodic contour will eventually dominate the results. It is seen that the Bach piece exhibits a much higher fractal dimension (D=1.58) than the Leksand tune (D=1.36). Indeed, there is a tendency in all our data (Table 1) that classical music has a higher fractal dimension than folk music tunes. However, a problem is that crossover effects are more apparent in the smaller data sets of the folk tunes (Figure 2b). The Table also lists characteristic exponents, α, of Levy-stable distributions, that were fitted to the distribution of melodic intervals, from our previous work [16]. It is observed that high fractal dimensions are correlated with low Lévy exponents.

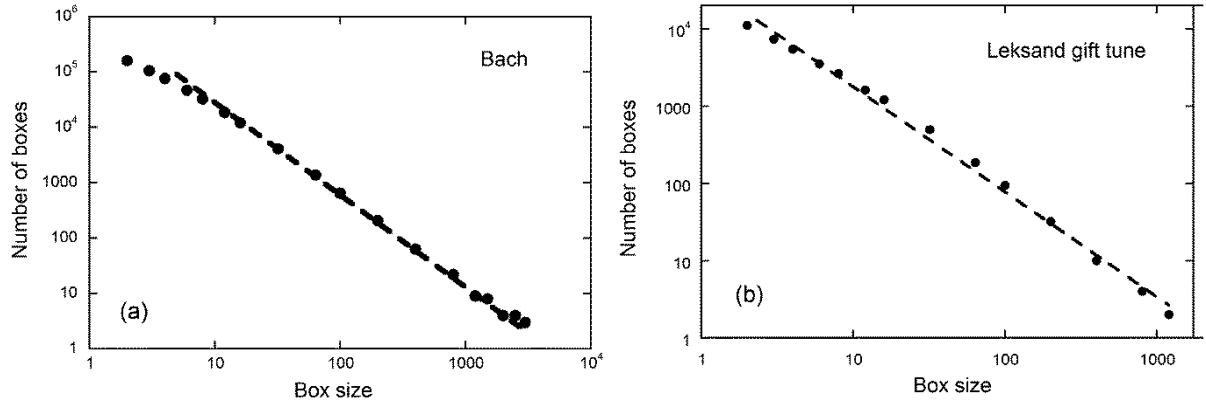

Fig. 2. Number of boxes as a function of box size for (a) Bach's Concerto for two violins in D minor, movement 2, and (b) Leksand gift tune. The dashed lines denote fits to the data points and the slope gives the fractal dimension.

**Table 1.** Number of data points, *N*, as well as fractal dimension, *D*, obtained by box counting and by DFA methods for the music pieces considered in this paper. Values from the DFA0 method were used for classical music and DFA1 method for folk music. In addition, the characteristic exponent, $\alpha$, of Levy-stable distributions that were fitted to the distribution of melodic intervals [16], are given for comparison. Uncertainties in *D* are estimated to be of the order of ± 0.03, while error bars of $\alpha$ are approximately ±0.10 for the classical pieces and ±0.20 for the folk tunes.

|  | N | D (box) | D(DFA) | $\alpha$ |
|---|---|---|---|---|
| Classical music: |  |  |  |  |
| Bach: Concerto for two violins in D minor | 1312 | 1.58 | 1.83 | 1.08 |
| Vivaldi: Concerto in A Minor 1st Movement | 1280 | 1.56 | 1.80 | 1.29 |
| Küchler: Concertino in D major, Op. 15 | 1040 | 1.55 | 1.72 | 1.37 |
| Folk music: |  |  |  |  |
| Gärdeby tune | 256 | 1.41 | 1.79 | 1.51 |
| Leksand "gift tune" | 256 | 1.36 | 1.50 | 1.56 |
| Lana-Ville's schottis | 160 | 1.33 | 1.50 | 1.73 |

Figure 3 shows the RMS fluctuations as a function of time interval for two classical musical scores. These calculations using the DFA0 algorithm show a very good scaling behavior over the whole range of bin sizes, with fractal dimensions of 1.80 and 1.83 . For classical music pieces applying detrending did not improve the scaling; instead the scaling range was slightly restricted. Hence it seems that there is no underlying linear or quadratic trend in these datasets. The situation is different for folk music, though, and in at least two cases linear detrending improved the accuracy of determining the scaling exponent. Figure 4 shows calculations by the DFA0 and DFA1 algorithms for the Leksand piece. There is a scaling range for short time intervals that crosses over to a rather constant behaviour for large time intervals. However, the scatter of the data points in significantly larger for the DFA0 method. Therefore the DFA1 data was used to compute *D* for the folk music tunes, giving acceptable scaling ranges extending almost decade in bin size. Fig. 4 thus illustrates one advantage of detrending.

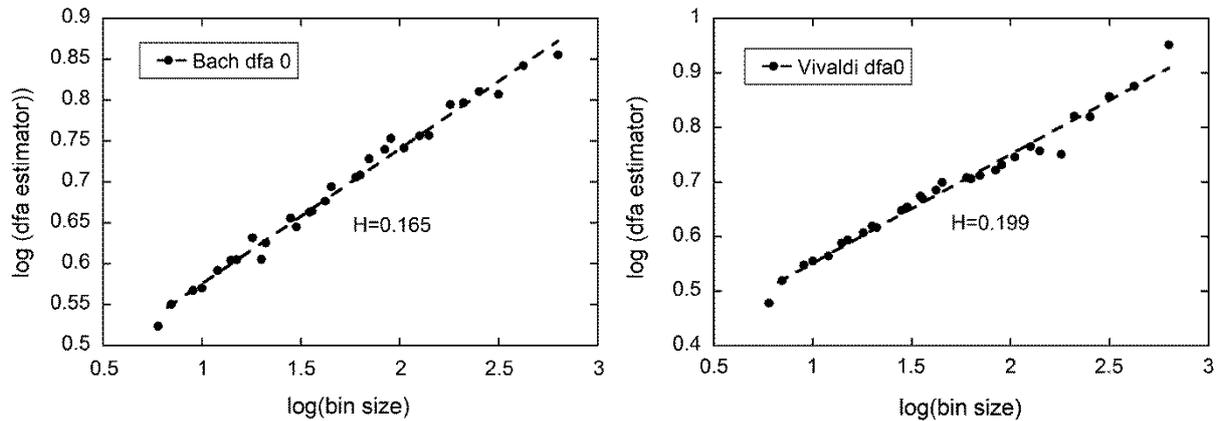

Fig. 3. Logarithm (base 10) of DFA estimator as a function of logarithm of length of time interval (bin size) for (a) Bach's Concerto for two violins in D minor, movement 2, and (b) Vivaldi's Concerto in A Minor 1st Movement. The lines are fits to the scaling ranges in the plots.

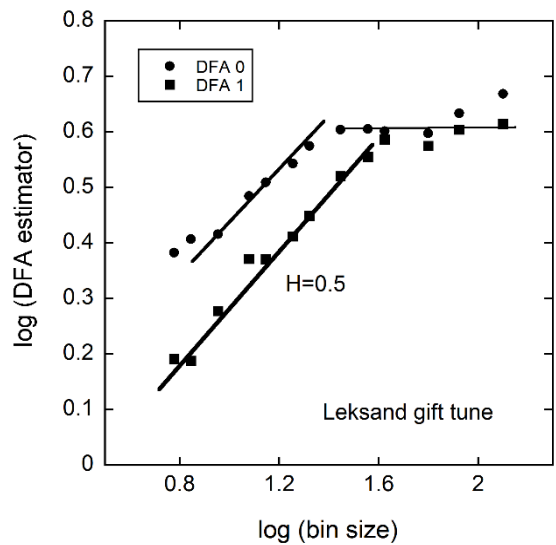

Fig. 4. Logarithm (base 10) of DFA estimator as a function of logarithm of length of time interval (bin size) for the Leksand gift tune, computed by the DAF0 and DFA1 methods. Full lines are fits to the scaling ranges as well as to the horizontal trend for sizes larger than the crossover.

The slope of the scaling range gives Hurst exponents H=0.18 and 0.50 for the Bach and Leksand pieces, respectively, equivalent to fractal dimensions of 1.82 and 1.50, respectively. Hence fractal dimensions obtained by DFA are consistently higher than box counting dimensions. These differences merit a thorough discussion. First it must be remembered that box counting computes D of the melodic contour, in which the pitch values are joined by straight lines, while DFA gives the dimension of a time series of pitch values. Secondly, it is known that box counting underestimates D close to the dimension of the embedding space (in our case 2) [34]. Thirdly, we are using relatively short time series in the present work. Hence the accuracy of various methods in this case is of prime importance.

Detrended Fluctuation Analysis (DFA) has been found to perform better than a variety of other methods in comparisons using synthetic data [30,35]. In particular, it is accurate for short fractional Brownian motion time series of up to 2048 data points [35]. Our present state of knowledge indicates that DFA should be the preferred method to analyze musical contours, although we stress that it needs to be validated also for fractional Lévy motions. The lower dimensions obtained by box counting may be due partly to the low accuracy of box counting for small values of $H$ and partly to a bias introduced by the image analysis of the contour lines connecting the points in the "music walk".

We now compare our fractal dimensions with previous studies, focusing on those that have used similar methods. Analysis of audio signals have yielded fractal dimensions in the range 1.6 – 1.7 from box counting [12] and 1.7 – 1.9 from variance methods [8]. A variation of $D$ between different music genres was found in the more extensive of these studies [8]. The RMS fluctuation method (DFA0) has been used mostly for classical music. Su and Wu [5] obtained fractal dimensions in the range 1.62-1.78 by analyzing the melodic contours of a number of classical pieces. However, their results showed strong crossover effects in most cases and probably more accurate values could have been obtained with the DFA1 method. Despite the methodical differences and uncertainties, literature values are generally in acceptable agreement with our results for classical music. It seems that classical music exhibits fractal dimensions in the range 1.6 – 1.9 and that folk music tunes, at least in some cases, would appear to have lower fractal dimensions. However, folk tunes are shorter, which results in small data sets, and the fractal dimensions may be significantly affected by crossover effects and cutoffs to the fractal range. In general, our results should be validated by analysis of a larger corpus of music. In that case, automated analysis, for example by using the MIDI Toolbox of the University of Jyväskylä [36], which computes melodic contours for monophonic MIDI files, will be necessary.

4. Conclusion

We have shown that melodic contours and time series of pitches exhibit a self-affine fractal scaling range. Fractal dimensions were computed by box counting and detrended fluctuation analysis. Our results confirm the trends obtained in a previous comparison of fractal analysis methods [35]. In particular, we infer that the DFA method is the best one, with the analyzed data showing wide scaling ranges with minor crossover effects. We observe that folk music tunes exhibit lower fractal dimensions than classical music, but the importance of crossover effects needs to be more accurately assessed in these cases. Recent results indicate that music can be modelled as a Levy motion [16]. Hence the accuracy of variance-based methods like DFA need to be re-examined for time series whose increments can be approximated by Levy-stable distributions. However, the present study substantiates the use of the DFA method in [16].